\documentclass[%
twocolumn,
showpacs,
preprintnumbers,
 amsmath,amssymb,
 aps,
 prc,
]{revtex4-1}

\usepackage{graphicx}
\usepackage{dcolumn}
\usepackage{bm}
\usepackage{lineno}
\begin{document}

\title{Selection of projectiles for producing trans-uranium nuclei in transfer reactions within an improved dinuclear system model}



\author{Long Zhu}


\affiliation{%
Sino-French Institute of Nuclear Engineering and Technology, Sun Yat-sen University, Zhuhai 519082, China\\
}%


\date{\today}

\begin{abstract}
 The multinucleon transfer process is one promising approach for producing neutron-rich heavy nuclei. The favorable projectile-target combinations play a key role in experiments. As one important approach to investigate the multinucleon transfer process, one new version of the dinuclear system (DNS) model is developed. The improved version of the DNS model is shown in detail. Based on deformation degree of freedom in the improved DNS model, one way of calculating excitation energies of primary fragments is proposed. And the remarkable improvements for describing experimental data of producing trans-uranium nuclei is noticed. To produce trans-uranium nuclei, the collisions of $^{48}$Ca, $^{136}$Xe, and $^{238}$U projectiles with the $^{238}$U target are investigated within the improved DNS model. Based on the potential energy surface, the influence of projectiles on the probabilities of transferring neutrons and protons from the projectiles to the target are studied. One behavior is found that the transfer probabilities strongly depend on the projectiles in the neutron stripping process, while relatively weak projectile dependence is noticed in the proton transferring process. The $^{136}$Xe and $^{238}$U projectiles show great advantages of cross sections for producing neutron-rich trans-uranium nuclei, although the fission probabilities are large.

\end{abstract}


\maketitle

\section{\label{int}INTRODUCTION}
 In the past decades, capture process of light particles or heavy ion induced fuion reactions play important roles for producing actinide and transactinide nuclei, including superheavy nuclei (SHN) \cite{Thoennessen01}. As well known, in the fusion reactions with stable combinations, the neutron-deficient heavy nuclei can mainly be produced. The radioactive beams induced fusion reactions can be used for producing the neutron-rich nuclei. However, based on present radioactive beam facilities, the beam intensities are much lower than the stable ones, which results in very low production rates. The multinucleon transfer (MNT) process has been proposed as one promising approach for producing neutron-rich heavy isotopes \cite{Volkov01,Zhang01,Loveland01,Watanabe01}, including the more neutron-rich SHN, which is one of important scientific goals in High Intensity heavy ion Accelerator Facility (HIAF) \cite{Yang01}.

 The MNT reactions for producing actinide nuclei were studied 40 years ago \cite{Hildenbrand01,Schadel01,Freiesleben01,Schadel02,Welch01,Mayer01,Moody01}. Recently, Wuenschel \emph{et al}. tried to produce SHN based on MNT process \cite{Wuenschel01}. Some promising results were presented. In Ref. \cite{Moody01}, the bombardments of $^{248}$Cm target with different projectiles were investigated. The Coulomb potential and stabilizing effect of the reaction Q values were proposed to understand the systematic trend of production cross sections in transfer reactions. The actinide projectiles provide high nucleon flux, which usually results in high cross sections of primary exotic nuclei. In Ref. \cite{Zagrebaev01}, from calculations of the mass and charge distributions of primary fragments, the advantages of cross sections in $^{238}$U induced MNT reactions are clearly shown. Nevertheless, the actinide nuclei are not as bounded as light ones. Therefore, the excitation energies of transferred products could be very high and the survival probabilities would be strongly suppressed. To find the favorable projectiles for producing neutron-rich heavy nuclei, it is worth to give investigations about influence of projectiles on the cross sections of the MNT products. The aim of this paper is to reveal the projectile effects on the cross sections based on the potential energy surface (PES) and select the favorable projectile for producing trans-uranium nuclei.

Several theoretical models, such as a Langevin-type approach \cite{Zagrebaev02,Zagrebaev05,Karpov01}, improved quantum molecular dynamics (ImQMD) model \cite{Tian01,Zhao01,Li02,Wang02}, time-dependent Hartree-Fock approach (TDHF) \cite{Sekizawa01,Jiang01}, GRAZING \cite{Winther01,Wen01}, and dinuclear system (DNS) model \cite{Adamian01,Penionzhkevich01,Feng01,Zhu07,Bao02,Zhu04}, have been applied to study the mechanism of MNT reactions. As one important method, one new version of the DNS model was developed, in which the temperature dependence of shell corrections on PES and the deformation degree of freedom in master equation were taken into account \cite{Zhu09,Zhu08}. In order to distinguish with other versions of DNS model, we call this version as ``DNS-sysu". In present paper, the DNS-sysu model is developed and applied to study the collisions of different projectiles with the $^{238}$U target for producing trans-target nuclei.

\section{\label{model}Theoretical framework}

The most important developments in the DNS-sysu in comparison to the previous version \cite{Zhu11,Feng02,Zhu10} are presented as follows. (i) The deformation degree of freedom is included in the master equation self-consistently. (ii) The temperature dependence of structure effects are considered in the model. (iii) The excitation energies of the primary fragments are calculated with consideration of deformation evolution.
(iv) One code for describing cooling process by using Monte Carlo method is developed.

In the DNS-sysu, the master equation can be written as \cite{Zhu08}
\begin{flalign}
\begin{split}\label{master}
&\frac{dP(Z_{1},N_{1},\beta_{2},t)}{dt}\\
&=\sum_{Z_{1}^{'}}W_{Z_{1},N_{1},\beta_{2};Z_{1}^{'},N_{1},\beta_{2}}(t)[d_{Z_{1},N_{1},\beta_{2}}P(Z_{1}^{'},N_{1},\beta_{2},t)\\
&-d_{Z_{1}^{'},N_{1},\beta_{2}}P(Z_{1},N_{1},\beta_{2},t)]\\
&+\sum_{N_{1}^{'}}W_{Z_{1},N_{1},\beta_{2};Z_{1},N_{1}^{'},\beta_{2}}(t)[d_{Z_{1},N_{1},\beta_{2}}P(Z_{1},N_{1}^{'},\beta_{2},t)\\
&-d_{Z_{1},N_{1}^{'},\beta_{2}}P(Z_{1},N_{1},\beta_{2},t)]\\
&+\sum_{\beta_{2}^{'}}W_{Z_{1},N_{1},\beta_{2};Z_{1},N_{1},\beta_{2}^{'}}(t)[d_{Z_{1},N_{1},\beta_{2}}P(Z_{1},N_{1},\beta_{2}^{'},t)\\
&-d_{Z_{1},N_{1},\beta_{2}^{'}}P(Z_{1},N_{1},\beta_{2},t)],
\end{split}
\end{flalign}
where $P(Z_{1},N_{1},\beta_{2},t)$ is the distribution probability for the fragment 1 with proton number $Z_{1}$ and neutron number $N_{1}$ at time $t$. $\beta_{2}$ is the dynamical deformation parameter of the DNS. $W_{Z_{1},N_{1},\beta_{2};Z_{1}^{'},N_{1},\beta_{2}}$, $W_{Z_{1},N_{1},\beta_{2};Z_{1},N_{1}^{'},\beta_{2}}$, and $W_{Z_{1},N_{1},\beta_{2};Z_{1},N_{1},\beta_{2}^{'}}$ denote the mean transition probabilities from the channels ($Z_{1}$, $N_{1}$, $\beta_{2}$) to ($Z_{1}^{'}$, $N_{1}$, $\beta_{2}$), ($Z_{1}$, $N_{1}$, $\beta_{2}$) to ($Z_{1}$, $N_{1}^{'}$, $\beta_{2}$), and ($Z_{1}$, $N_{1}$, $\beta_{2}$) to ($Z_{1}$, $N_{1}$, $\beta_{2}^{'}$), respectively.
 $d_{Z_{1},N_{1},\beta_{2}}$ is
the microscopic dimension (the number of channels) corresponding to the macroscopic state ($Z_{1}$, $N_{1}$, $\beta_{2}$) \cite{Norenberg01}. For the degrees of freedom of charge and neutron number, the sum is taken over all possible proton and neutron numbers that fragment 1 may take, but only one nucleon transfer is considered in the model ($Z_{1}^{'}=Z_{1}\pm1$; $N_{1}^{'}=N_{1}\pm1$). For $\beta_{2}$, we take the range of -$0.5 \sim 0.5$. The evolution step length is 0.01. The transition probability is related to the local excitation energy \cite{Zhu08,Ayik01}.

The PES is defined as
\begin{flalign}
\begin{split}
 U(Z_{1}, &N_{1}, \beta_{2}, R_{\textrm{cont}})=\Delta(Z_{1}, N_{1})+\Delta(Z_{2}, N_{2})\\
& +V_{\textrm{cont}}(Z_{1}, N_{1}, \beta_{2}, R_{\textrm{cont}})+\frac{1}{2}C_{1}(\beta_{2}^{1}-\beta_{2}^{\textrm{p}})^{2}\\
&+\frac{1}{2}C_{2}(\beta_{2}^{2}-\beta_{2}^{\textrm{t}})^{2}.
\end{split}
\end{flalign}
Here, $\Delta(Z_{i}, N_{i})$ ($i=1$, 2) is mass excess of the fragment $i$, including the paring and shell corrections, which can be written as \cite{Zhu09}
\begin{flalign}
\begin{split}\label{correc}
\Delta(Z_{i},& N_{i})=Z_{i}\Delta(^{1}H)+N_{i}\Delta(n)-a_{\textrm{v}}(1-\kappa I^{2})A_{i}\\
&+a_{\textrm{s}}(1-\kappa I^{2})A_{i}^{2/3}+a_{\textrm{c}}Z_{i}^{2}A_{i}^{-1/3}-c_{4}Z_{i}^{2}A_{i}^{-1}\\
&-E_{\textrm{pair}}(Z_{i}, N_{i})+E_{\textrm{sh}}(Z_{i}, N_{i}).
\end{split}
\end{flalign}
The liquid drop parameters can be seen in Ref. \cite{Samaddar01,Zhu09}. The pairing energy $E_{\textrm{pair}}(Z_{i}, N_{i})=E_{\textrm{pair}}^{0}(Z_{i}, N_{i})e^{-(E^{\prime}/a)^{2}}$. The shell correction energy $E_{\textrm{sh}}(Z_{i}, N_{i})=E_{\textrm{sh}}^{0}(Z_{i}, N_{i})e^{-E^{\prime}/E_{\textrm{d}}}$. $E_{\textrm{d}}=5.48A_{i}^{1/3}/(1+1.3A_{i}^{-1/3})$ MeV. $a=A/12$ MeV$^{-1}$. $E_{\textrm{pair}}^{0}$ is the pairing energy on ground state, which is given by \cite{Zhu09}
 \begin{eqnarray}\label{pair}
    E_{\textrm{pair}}^{0}=
    \left\{
    \begin{aligned}
     &2a_{\textrm{p}}/A_{i}^{1/2},  &{\rm for\;even\;}Z, {\rm even\;}N\;{\rm nuclei};\\
     &a_{\textrm{p}}/A_{i}^{1/2},   &{\rm for\;odd\;}A\;{\rm nuclei};\\
     &0,  &{\rm for\;odd\;}Z, {\rm odd\;}N\;{\rm nuclei}.\\
    \end{aligned}
    \right.
  \end{eqnarray}
Here, $a_{\textrm{p}}$ equals 12 MeV. $E^{\prime}=E_{\textrm{diss}}\times A_{i}/A_{\textrm{tot}}$. $A_{\textrm{tot}}$ is the total mass number of the system. Other liquid drop terms weakly depend on the fragments temperature. $E_{\textrm{diss}}$ can be calculated as
\begin{flalign}
\begin{split}
E_{\textrm{diss}}(J,t)=&E_{\textrm{c.m.}}-V_{\textrm{cont}}(Z_{\textrm{p}}, N_{\textrm{p}},\beta_{2}, R_{\textrm{cont}})\\
&-\frac{(J^{'}(t)\hbar)^{2}}{2\zeta_{\textrm{rel}}}-E_{\textrm{rad}}(J,t),
\end{split}
\end{flalign}
where, $J^{'}(t)$ ($=J_{\textrm{st}}+(J-J_{\textrm{st}})e^{-t/\tau_{J}}$) is the relative angular momentum at time $t$. $J$ is initial entrance angular momentum.  $J_{\textrm{st}}=\frac{\zeta_{\textrm{rel}}}{\zeta_{\textrm{tot}}}J$. $\zeta_{\textrm{rel}}$ and $\zeta_{\textrm{tot}}$ are the relative and total moments of inertia, respectively. $\tau_{J}=12\times10^{-22}$ s.
$E_{\textrm{rad}}(J,t)=[E_{\textrm{c.m.}}-V_{\textrm{cont}}(Z_{1}, N_{1},\beta_{2}, R_{\textrm{cont}})-\frac{(J\hbar)^{2}}{2\zeta_{\textrm{rel}}}]e^{-t/\tau_{R}}$. $\tau_{R}$ ($=2\times10^{-22}$ s) is the characteristic relaxation time of radial energy.

$\beta_{2}^{1}$ ($=\beta_{2}^{\textrm{p}}+\delta \beta_{2}^{1}$) and $\beta_{2}^{2}$ ($=\beta_{2}^{\textrm{t}}+\delta \beta_{2}^{2}$) are quadrupole deformation parameters of projectile-like fragment (PLF) and target-like fragment (TLF), respectively. $\beta_{2}^{\textrm{p}}$ and $\beta_{2}^{\textrm{t}}$ are static deformation parameters of projectile and target, respectively, which are taken from Ref. \cite{Moller01}. $C_{1}\delta \beta_{2}^{1}=C_{2}\delta \beta_{2}^{2}$. $\delta \beta_{2}^{1}+ \delta \beta_{2}^{2}=2\beta_{2}$. $C_{1,2}$ are the liquid drop model stiffness parameters of the fragments, the description of which can be seen in Ref. \cite{Myers01}.

The effective nucleus-nucleus interaction potential $V_{\textrm{cont}}(Z_{1}, N_{1},\beta_{2}, R_{\textrm{cont}})$ between fragments 1 and 2 can be written as
\begin{flalign}
\begin{split}
 V_{\textrm{cont}}(Z_{1}, N_{1},\beta_{2}, R_{\textrm{cont}})&=V_{\textrm{N}}(Z_{1}, N_{1},\beta_{2}, R_{\textrm{cont}})+\\
&V_{\textrm{C}}(Z_{1}, N_{1},\beta_{2}, R_{\textrm{cont}}).
\end{split}
\end{flalign}
For collision systems with potential pocket in the entrance channel, the potential energy is calculated at the bottom of potential pocket. For the reactions with no potential pockets, the position where the nucleon transfer process takes place can be obtained with the equation: $R_{\textrm{cont}}=R_{1}(1+\beta_{2}^{1}Y_{20}(\theta_{1}))+R_{2}(1+\beta_{2}^{2}Y_{20}(\theta_{2}))+0.7$ fm \cite{Zhu07}. Here, $R_{1,2}=1.16A_{1,2}^{1/3}$. $\theta_{1}=\theta_{2}=0$. The detailed description of nuclear potential and Coulomb potential can be seen in Refs. \cite{Zhu09,Wong01}.

The local excitation energy of the DNS is determined by
\begin{flalign}
\begin{split}
E^{*}_{\textrm{DNS}}(Z_{1},N_{1},\beta_{2},J,t)=&E_{\textrm{diss}}(J,t)-
[U(Z_{1}, N_{1}, \beta_{2}, R_{\textrm{cont}})\\
&-U(Z_{\textrm{p}}, N_{\textrm{p}},\beta_{2}, R_{\textrm{cont}})].
\end{split}
\end{flalign}

In the previous DNS code, the excitation energy of primary products ($Z_{i}$, $N_{i}$) with incident angular momentum $J$ is calculated by
\begin{flalign}
\begin{split}\label{exip}
E^{*}= E^{*}_{\textrm{DNS}}(Z_{i},N_{i},J,t=\tau_{\textrm{int}})\frac{Z_{i}+N_{i}}{A_{\textrm{tot}}}.
\end{split}
\end{flalign}
$A_{\textrm{tot}}$ is the total mass number of the system. $\tau_{\textrm{int}}$ is the interaction time of the collision system.
In the DNS-sysu, with consideration of the deformation evolution, the excitation energy of primary products can be calculated with following equation.
\begin{flalign}
\begin{split}\label{exi}
&E^{*}=\\
&\frac{\sum\limits_{\beta_{2}}[ P(Z_{i},N_{i},\beta_{2},t=\tau_{\textrm{int}}) E^{*}_{\textrm{DNS}}(Z_{i},N_{i},\beta_{2},J,t=\tau_{\textrm{int}})]}{\sum\limits_{\beta_{2}} P(Z_{i},N_{i},\beta_{2},t=\tau_{\textrm{int}})}\\
&\times\frac{Z_{i}+N_{i}}{A_{\textrm{tot}}}.
\end{split}
\end{flalign}
Here, it is worth to mention that $E^{*}$ is different from $E^{\prime}$ described in Eq. (\ref{correc}).

The production cross sections of the primary products in transfer reactions can be calculated as follows:
\begin{flalign}
\begin{split}\label{cros}
 \sigma_{\textrm{pr}}&(Z_{1},N_{1},E_{\textrm{c.m.}})=\frac{\pi\hbar^{2}}{2\mu E_{\textrm{c.m.}}}\sum_{J=0}^{J_{\textrm{max}}}(2J+1) \\
& \times T_{\textrm{cap}}(J,E_{\textrm{c.m.}})\times\sum_{\beta_{2}} P(Z_{1},N_{1},\beta_{2},E_{\textrm{c.m.}}).
\end{split}
\end{flalign}
Here, the second sum is taken over all possible $\beta_{2}$ that may take. The transmission probability can be written as \cite{Hill01}
\begin{flalign}
\begin{split}\label{}
&T_{\textrm{cap}}(J,E_{\textrm{c.m.}})= \\
&\frac{1}{1+\textrm{exp}\{-\frac{2\pi}{\hbar\omega(J)}[E_{\textrm{c.m.}}-B-\frac{\hbar^{2}}{2\mu R_{B}^{2}(J)}J(J+1)]\}},
\end{split}
\end{flalign}
where $\hbar\omega(J) = \hbar \sqrt{-\frac{1}{\mu}\frac{\partial^{2} V}{\partial r^{2}}}\Big | _{R=R_{B}}$ is the width of the parabolic Coulomb barrier at the position $R_{B}(J)$. For the reactions without potential pockets in the entrance channel (there are no ordinary barriers: the potential energies of these nuclei are everywhere repulsive) and the incident energies are above the interaction potentials at the contact configurations for different entrance angular momentum, the $T_{\textrm{cap}}$ is estimated as 1.
\begin{figure}
\begin{center}
\includegraphics[width=8.8cm,angle=0]{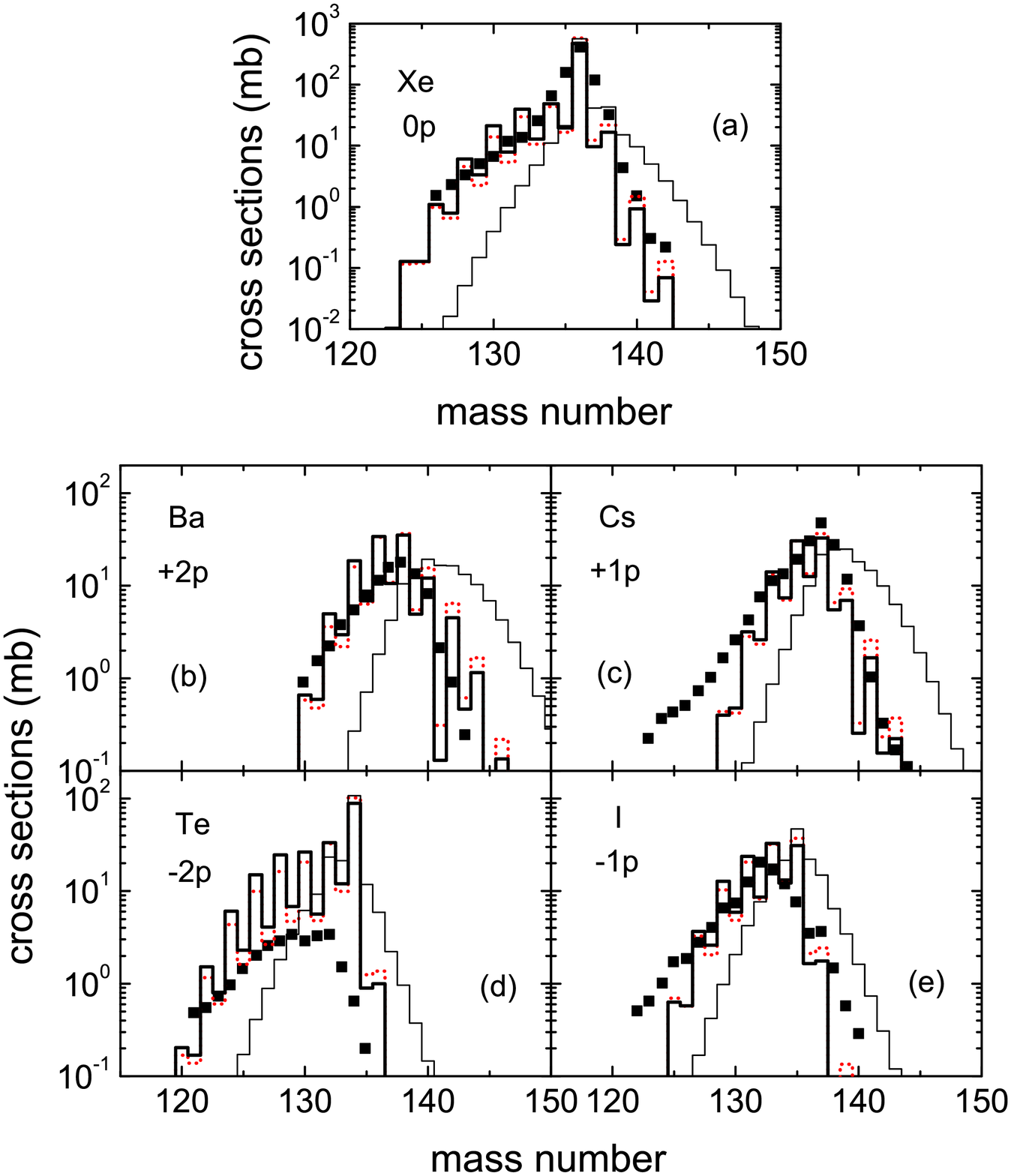}
\caption{\label{exp} Isotope distributions of projectile-like products in the reaction $^{136}$Xe + $^{238}$U. The experimental data \cite{Vogt01} are denoted with solid squares. The incident energy $E_{\textrm{c.m.}}$=636 MeV. The thick and thin solid lines denote the results of final products and primary products, respectively, calculated in the DNS-sysu. The dotted lines denote the calculation results for excitation energies calculated by Eq. (\ref{exip}).}
\end{center}
\end{figure}

\begin{figure*}
\begin{center}
\includegraphics[width=15cm,angle=0]{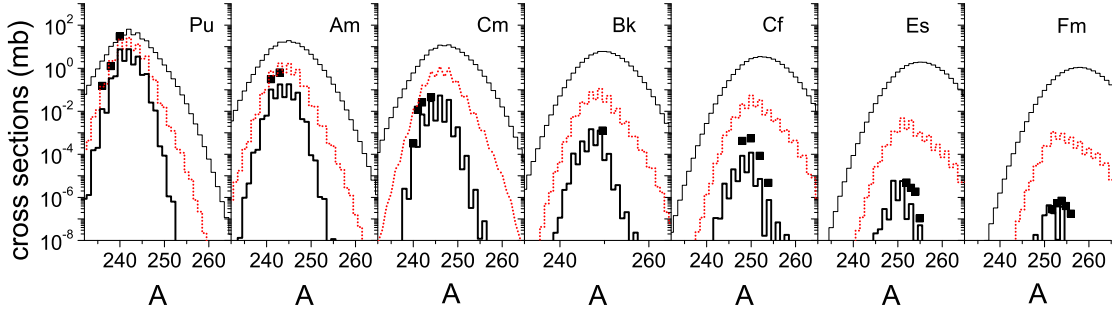}
\caption{\label{exp2} Comparison of experimental data with the calculated production cross sections of trans-uranium nuclei in the reaction $^{238}$U + $^{238}$U ($E=7.5$ MeV/u). The experimental data \cite{Schadel01} are denoted with solid squares. The thin and thick solid lines denote the results of primary products and final products, respectively, calculated in the DNS-sysu. The dotted lines denote the calculation results for excitation energies calculated by Eq. (\ref{exip}).}
\end{center}
\end{figure*}

The cooling process of excitation fragments is calculated with statistical model. The excitation energies of the primary fragments are calculated in the Eq. (\ref{exi}). The Monte Carlo method is used to obtain the probabilities of all possible decay channels. The decay chain is ended when fission happens or the fragments reach the ground state. In the $\emph{i}$th de-excitation step the probability of the $\emph{s}$ event can be written as
\begin{flalign}
\begin{split}
P_{s}(E^{*}_{i})=\frac{\Gamma_{s}(E^{*}_{i})}{\Gamma_{\textrm{tot}}(E^{*}_{i})},
\end{split}
\end{flalign}
where, $s=$ n, p, $\alpha$, $\gamma$, and fission. $E_{i}^{*}$ is the excitation
energy before $i$th decay step, which can be calculated from the equation $E^{*}_{i+1}=E^{*}_{i}-B_{i}$. $B_{i}$ is the separation energy of particle or energy taken by the $\gamma$ ray in the $i$th step. $\Gamma_{\textrm{tot}}=\Gamma_{\textrm{n}}+\Gamma_{\textrm{f}}+\Gamma_{\textrm{p}}+\Gamma_{\textrm{$\alpha$}}+\Gamma_{\textrm{$\gamma$}}$. The partial decay widths of the excited nucleus for the evaporation of the light particle $\nu=$(n, p, $\alpha$) can be estimated using the Weisskopf-Ewing theory \cite{Weisskopf02},
 \begin{flalign}
\begin{split} \label{decay_widths}
\Gamma_{\nu}&(E^{*},J)=\frac{(2s_{\nu}+1)m_{\nu}}{\pi^{2}\hbar^{2}\rho(E^{*},J)}\\
&\times\int_{I_{\nu}}\varepsilon \rho(E^{*}-B_{\nu}-\varepsilon,J)\sigma_{\textrm{inv}}(\varepsilon)\ d\varepsilon,
\end{split}
\end{flalign}
where $I_{\nu} = [0, E^{*}-B_{\nu}]$. $\sigma_{\textrm{inv}}$ is the inverse reaction cross section for particle $\nu$ with channel energy $\varepsilon$. The Coulomb barrier for charged particle emission is calculated as shown in Ref. \cite{Zubov01}.

The fission decay width is usually calculated within the Bohr-Wheeler (BW) transition-state method \cite{Bohr01}
 \begin{flalign}
\begin{split}
\Gamma&_{\textrm{f}}(E^{*},J)=\frac{1}{2\pi \rho_{\textrm{f}}(E^{*},J)}\\
&\times\int_{I_{\textrm{f}}}\frac{\rho_{\textrm{f}}(E^{*}-B_{\textrm{f}}-\varepsilon, J)\ d\varepsilon}{1+\textrm{exp}[-2\pi(E^{*}-B_{\textrm{f}}-\varepsilon)/\hbar\omega]},
\end{split}
\end{flalign}
where $I_{\textrm{f}} = [0, E^{*}-B_{\textrm{f}}]$. The temperature depedent fission barrier height is obtained by
 \begin{flalign}
\begin{split}
B_{\textrm{f}}(E^{*})=B_{\textrm{mac}}^{0}(1-x_{\textrm{ld}}T^{2})-E_{\textrm{sh}}^{0}e^{-E^{*}/E_{d}},
\end{split}
\end{flalign}
where, $B_{\textrm{mac}}^{0}$ is the macroscopic part of fission barrier on the ground state \cite{Ivanyuk01}. $T=\sqrt{E^{*}/a}$. The level density parameter $a=A/12$ MeV$^{-1}$. $x_{\textrm{ld}}$=0.04 \cite{Denisov01}.

The $\gamma$ emission width can be written as
 \begin{flalign}
\begin{split}
\Gamma_{\gamma}(E^{*},J)=\frac{3}{\rho(E^{*},J)}
\int_{0}^{E^{*}}\rho(E^{*}-\varepsilon, J)f_{E1}(\varepsilon)\ d\varepsilon.
\end{split}
\end{flalign}
Here, $f_{E1}$ is the strength function. Assuming that the the electric dipole radiation dominates in $\gamma$ emission. $f_{E1}$ is given by the following expression
 \begin{flalign}
\begin{split}
f_{E1}(\varepsilon)=\frac{4}{3\pi}\frac{1+\kappa}{mc^{2}}\frac{e^{2}}{\hbar c}\frac{ZN}{A}\frac{\Gamma_{G}\varepsilon^{4}}{(\Gamma_{G}\varepsilon)^{2}+(E^{2}_{G}-\varepsilon^{2})^{2}},
\end{split}
\end{flalign}
where, $\kappa=0.75$. $\Gamma_{G}$=5 MeV. $E_{G}$ is calculated as
 \begin{flalign}
\begin{split}
E_{G}=\frac{167.23}{A^{1/3}\sqrt{1.959+14.074A^{-1/3}}}.
\end{split}
\end{flalign}

The level density is taken as the standard Fermi-gas model \cite{Egidy01},
 \begin{flalign}
\begin{split}
\rho(E^{*},J)=&\frac{2J+1}{24\sqrt{2}\sigma^{3}a^{1/4}(E^{*}-\delta)^{5/4}}\\
&\times\textrm{exp}[2\sqrt{a(E^{*}-\delta)}-\frac{(J+1/2)^{2}}{2\sigma^{2}}],
\end{split}
\end{flalign}
where, $\delta$ is the shift energy as shown in Eq. (\ref{pair}). $\sigma^{2}=6\times0.24A^{2/3}\sqrt{a(E^{*}-\delta)}/\pi^{2}$

The first two improvements of the DNS-sysu model has been tested for producing heavy nuclei around $N=126$ \cite{Zhu08,Zhu09} and exotic nuclei near neutron drip line \cite{Zhu12} in combination with GEMINI code. Figure \ref{exp} shows the calculated production cross sections of Te, I, Xe, Cs, and Ba isotopes in the MNT reaction $^{136}$Xe + $^{238}$U within the DNS-sysu model. The experimental data \cite{Vogt01} are also shown. The calculations can reproduce the experimental data well. The calculated cross section of $^{134}$Te is higher than experimental data, which is due to overestimation of shell effects from $Z=50$ and $N=82$. It can be seen that the fission probabilities are small and several neutrons are evaporated in cooling process. The calculated production cross sections of final products based on the cooling of primary products with excitation energies calculated by Eq. (\ref{exip}) are also shown. No much difference is noticed between the two ways of calculating excitation energies.

The production of trans-uranium nuclei in collisions of actinide nuclei is calculated within the DNS-sysu model for the first time. The comparison of calculated production cross sections with the experimental data in the reaction $^{238}$U + $^{238}$U are presented in Fig. \ref{exp2}. A rather good agreement between calculated results and experimental data is shown. The remarkable improvement for reproducing experimental data is shown based on the excitation energies calculated with consideration of deformation evolution, which confirm the inclusion of deformation degree of freedom is reliable in the DNS-sysu. The DNS-sysu can be used for investigating the production of nuclei with large mass range in MNT reactions .

\section{\label{result}Results and discussion}
\begin{figure}
\begin{center}
\includegraphics[width=8.5cm,angle=0]{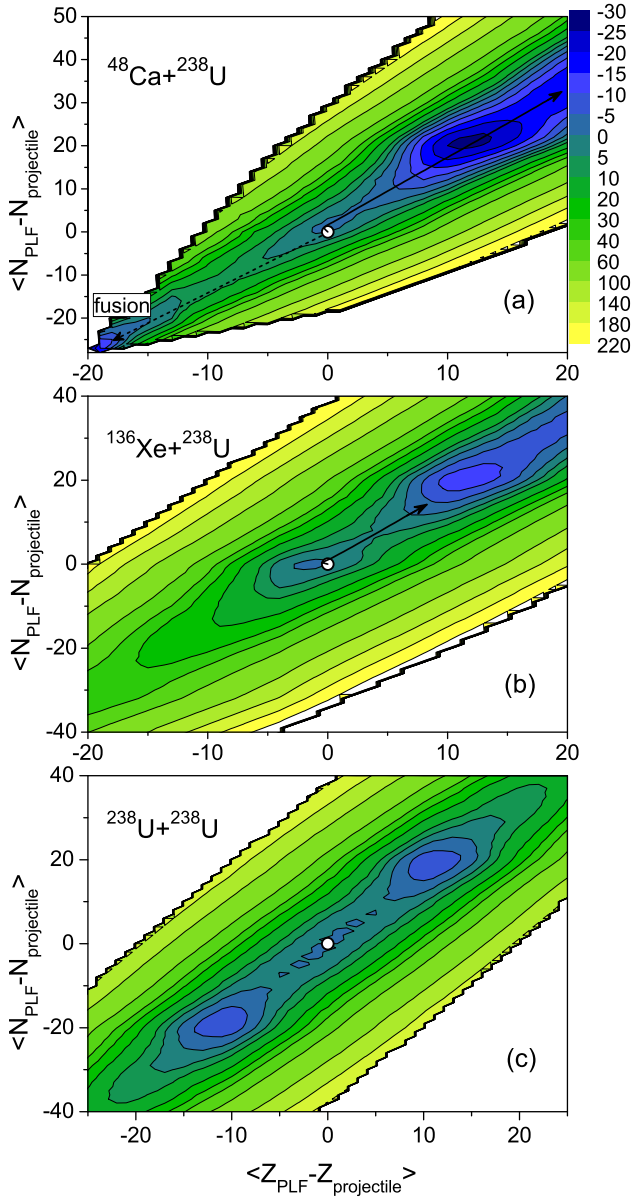}
\caption{\label{PES} Contour plots of the PES (in MeV) with the drift trajectories of the first moments of PLF distributions in a $\langle Z_{\textrm{PLF}}-Z_{\textrm{projectile}}\rangle$, $\langle N_{\textrm{PLF}}-N_{\textrm{projectile}}\rangle$ plane for the reactions $^{48}$Ca + $^{238}$U ($E_{\textrm{c.m.}}$= 201 MeV) (a), $^{136}$Xe + $^{238}$U ($E_{\textrm{c.m.}}$ = 537 MeV) (b), and $^{238}$U + $^{238}$U (c). The dashed arrow in (a) denote the pathway of fusion. The circles denote the injection points. }
\end{center}
\end{figure}
The collisions of the projectiles $^{48}$Ca, $^{136}$Xe, and $^{238}$U with the $^{238}$U target for producing transtarget nuclei are investigated in the DNS-sysu model. Figure \ref{PES}(a) shows the contours of PES with the drift trajectories of the first moments of PLF distributions for the reaction $^{48}$Ca + $^{238}$U. The dashed arrow shows the trajectory to fusion. In fusion process, the nucleons are transferred from $^{48}$Ca to $^{238}$U. One inner barrier need to be overcome to form a compound nucleus. Actually, the trend of neutron and proton pickup can be seen from the drift path of the first moments of PLF distributions. As discussed in Ref. \cite{Zhu07}, mass asymmetry relaxation plays an important role during the heavy ions collisions. Due to shell closures $Z=82$ and $N=126$, which corresponding to 10 protons and 20 neutrons transferring from $^{238}$U to $^{48}$Ca, one deep valley can be seen. It seems like that the shell effects are unfavorable for producing trans-uranium nuclei.

In comparison to the reaction $^{48}$Ca + $^{238}$U, the drift trajectory of the first moments of PLF distributions in the reaction $^{136}$Xe + $^{238}$U is much shorter, as shown in Fig. \ref{PES}(b). The shell effects on PES around the injection point are more obvious than that in the reaction $^{48}$Ca + $^{238}$U, which could strongly influence the nucleon transfer at low incident energies. For the symmetry configuration $^{238}$U + $^{238}$U, the production of trans-uranium nuclei is mainly due to fluctuation. Also, because of the shell closures $Z=82$ and $N=126$, one deep well can be seen, which could attract the system for producing trans-uranium nuclei, as shown in Fig. \ref{PES}(c). Based on the above discussions about PES, it can be easily make conjecture that the reaction $^{238}$U + $^{238}$U is more favorable for producing primary trans-uranium fragments.

\begin{figure}
\begin{center}
\includegraphics[width=8.5cm,angle=0]{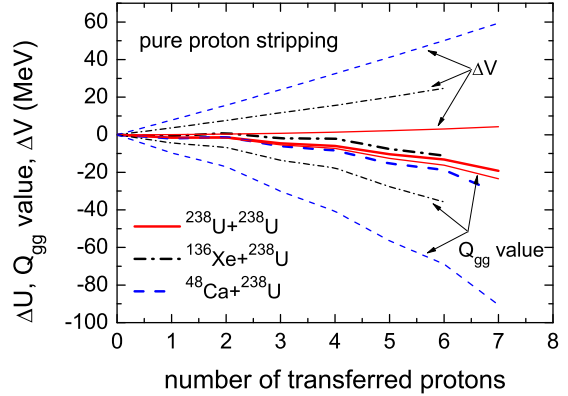}
\caption{\label{Vp} $\Delta U$(=$U(Z_{\textrm{P}},N_{\textrm{P}})-U(Z_{1},N_{1})$), Q$_{\textrm{gg}}$(=$M_{\textrm{P}}+M_{\textrm{T}}-M_{\textrm{PLF}}-M_{\textrm{TLF}}$), $\Delta V$ (=$V_{\textrm{cont}}(Z_{\textrm{P}},N_{\textrm{P}})-V_{\textrm{cont}}(Z_{\textrm{1}},N_{\textrm{1}})$) as functions of proton stripping number for the reactions $^{48}$Ca + $^{238}$U, $^{136}$Xe + $^{238}$U, and $^{238}$U + $^{238}$U.}
\end{center}
\end{figure}

\begin{figure}
\begin{center}
\includegraphics[width=8.5cm,angle=0]{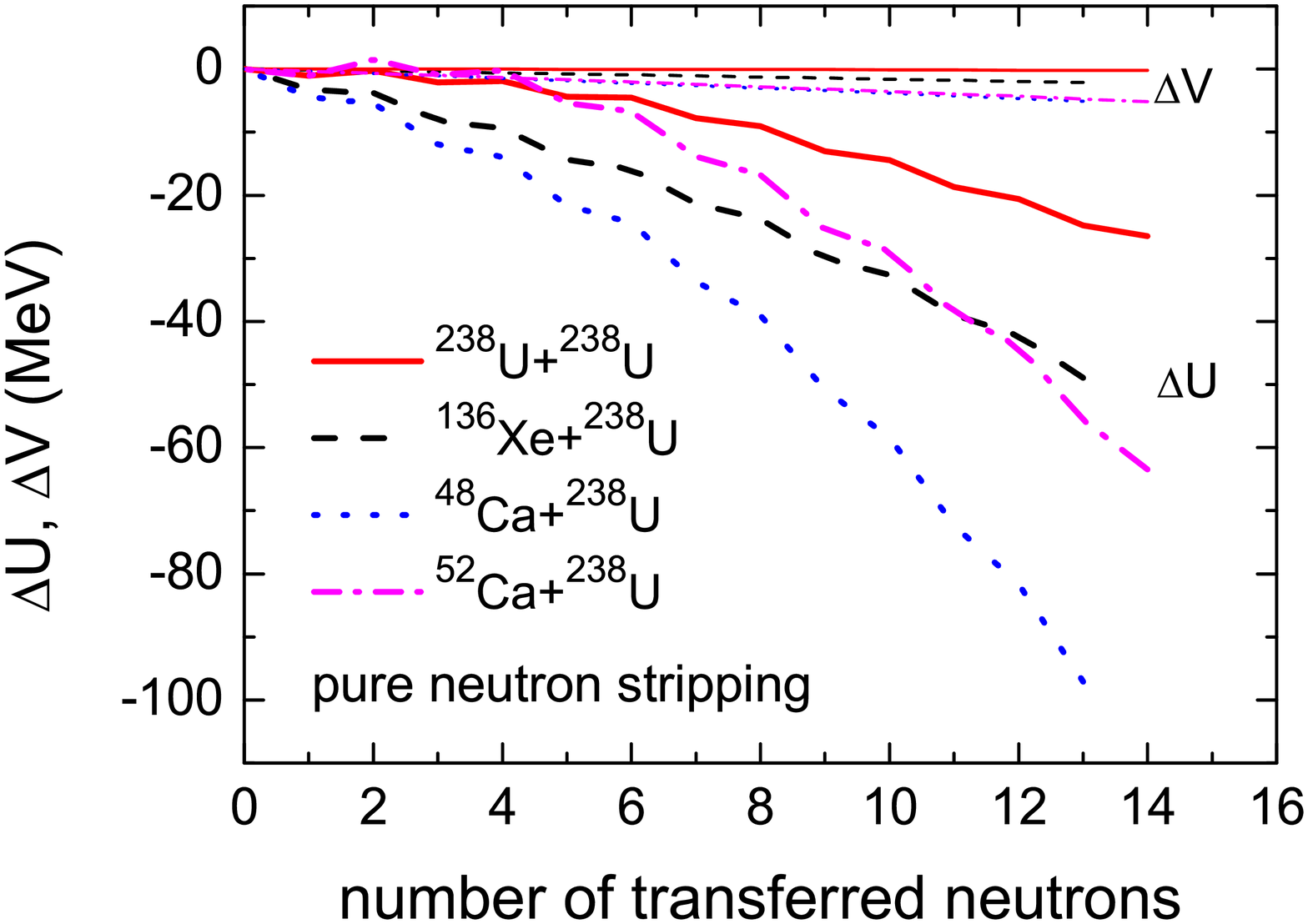}
\caption{\label{Vn} $\Delta U$ and $\Delta V$ as functions of neutron stripping number in the reactions $^{48}$Ca + $^{238}$U, $^{136}$Xe + $^{238}$U, $^{238}$U + $^{238}$U, and $^{52}$Ca + $^{238}$U.}
\end{center}
\end{figure}
In order to gain an insight into the influence of projectiles on PES, the values of $\Delta U$(=$U(Z_{\textrm{P}},N_{\textrm{P}})-U(Z_{1},N_{1})$), Q$_{\textrm{gg}}$(=$M_{\textrm{P}}+M_{\textrm{T}}-M_{\textrm{PLF}}-M_{\textrm{TLF}}$), $\Delta V$ (=$V_{\textrm{cont}}(Z_{\textrm{P}},N_{\textrm{P}})-V_{\textrm{cont}}(Z_{\textrm{1}},N_{\textrm{1}})$) for the reactions $^{48}$Ca + $^{238}$U, $^{136}$Xe + $^{238}$U, and $^{238}$U + $^{238}$U are extracted and shown in Fig. \ref{Vp}. It can be seen that the Q$_{\textrm{gg}}$ values are all negative for these reactions in pure proton stripping process and the absolute values increase with an increasing number of stripped protons from projectiles. Also, the Q$_{\textrm{gg}}$ (=$M_{\textrm{P}}+M_{\textrm{T}}-M_{\textrm{PLF}}-M_{\textrm{TLF}}$) value decreases much stronger for lighter projectile induced reaction. For $^{48}$Ca + $^{238}$U, during proton stripping process, much energy needs to be absorbed. However, the change of Coulomb potential $\Delta V$(=$V_{\textrm{cont}}(Z_{\textrm{P}},N_{\textrm{P}})-V_{\textrm{cont}}(Z_{\textrm{1}},N_{\textrm{1}})$) is very positive large in proton stripping process in the reaction $^{48}$Ca + $^{238}$U. The Coulomb potential drives system to more asymmetry configurations. The values of $\Delta V$ can compensate the negative Q$_{\textrm{gg}}$. We also show the results $\Delta U$, which is composed of Q$_{\textrm{gg}}$ and variation of Coulomb potential. The thick lines denote the results of $\Delta U$. One can see that the values of $\Delta U$ for these reactions are close. One the other hand, the probabilities of transferring protons from projectiles to the target would not show much difference for these reactions.

We show the case of pure neutron stripping in Fig. \ref{Vn}. Unlike the case of proton stripping, in neutron transfer process, the variations of Coulomb potentials are very small for the reactions $^{48}$Ca + $^{238}$U, $^{136}$Xe + $^{238}$U, $^{238}$U + $^{238}$U, and $^{52}$Ca + $^{238}$U. It can be seen that $\Delta U$ decreases strongly with the increasing number of transferred neutrons. Obviously, the high barriers need to be overcome in neutron stripping process, especially in the reaction $^{48}$Ca + $^{238}$U. Therefore, in comparison to the reaction $^{238}$U + $^{238}$U, the neutron transfer probabilities are strongly suppressed in the reaction $^{48}$Ca + $^{238}$U. The N/Z ratio of $^{48}$Ca is 1.4, which is much smaller than 1.59 of $^{238}$U. The effects of charge equilibration could be the reason for high barrier for transferring neutrons from $^{48}$Ca to the target. To clarify this effect, we show the results of the reaction $^{52}$Ca + $^{238}$U with a more neutron-rich projectile. One can see that the barrier for neutron stripping is also strongly suppressed in the reaction $^{52}$Ca + $^{238}$U. The projectile $^{238}$U still show great advantages in neutron stripping process.
We can make conclusion that influence of projectiles on PES in neutron transferring process is much stronger than that in proton transferring. Also, one conjecture can be made that the cross sections of transferring neutrons in the neutron stripping channel strongly depend on the projectiles. On the other side, the dependence of cross sections on projectiles in the proton stripping channel is relatively weak.
\begin{figure}
\begin{center}
\includegraphics[width=8.5cm,angle=0]{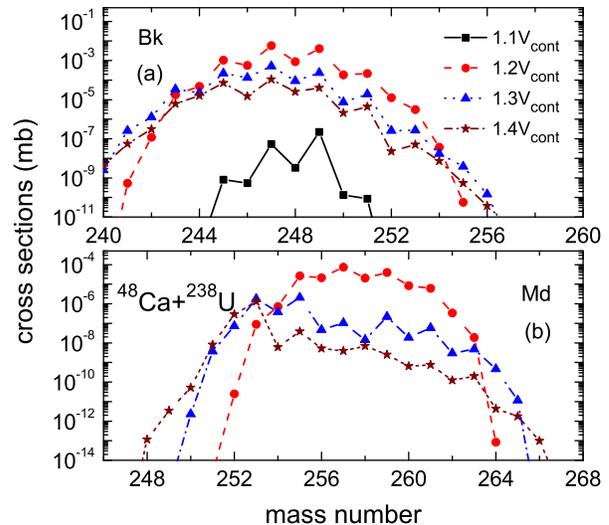}
\caption{\label{E48} Calculated cross sections of Bk (a) and Md (b) isotopes at different incident energies in the reaction $^{48}$Ca + $^{238}$U. V$_{\textrm{cont}}$=169 MeV. }
\end{center}
\end{figure}

\begin{figure}
\begin{center}
\includegraphics[width=8.5cm,angle=0]{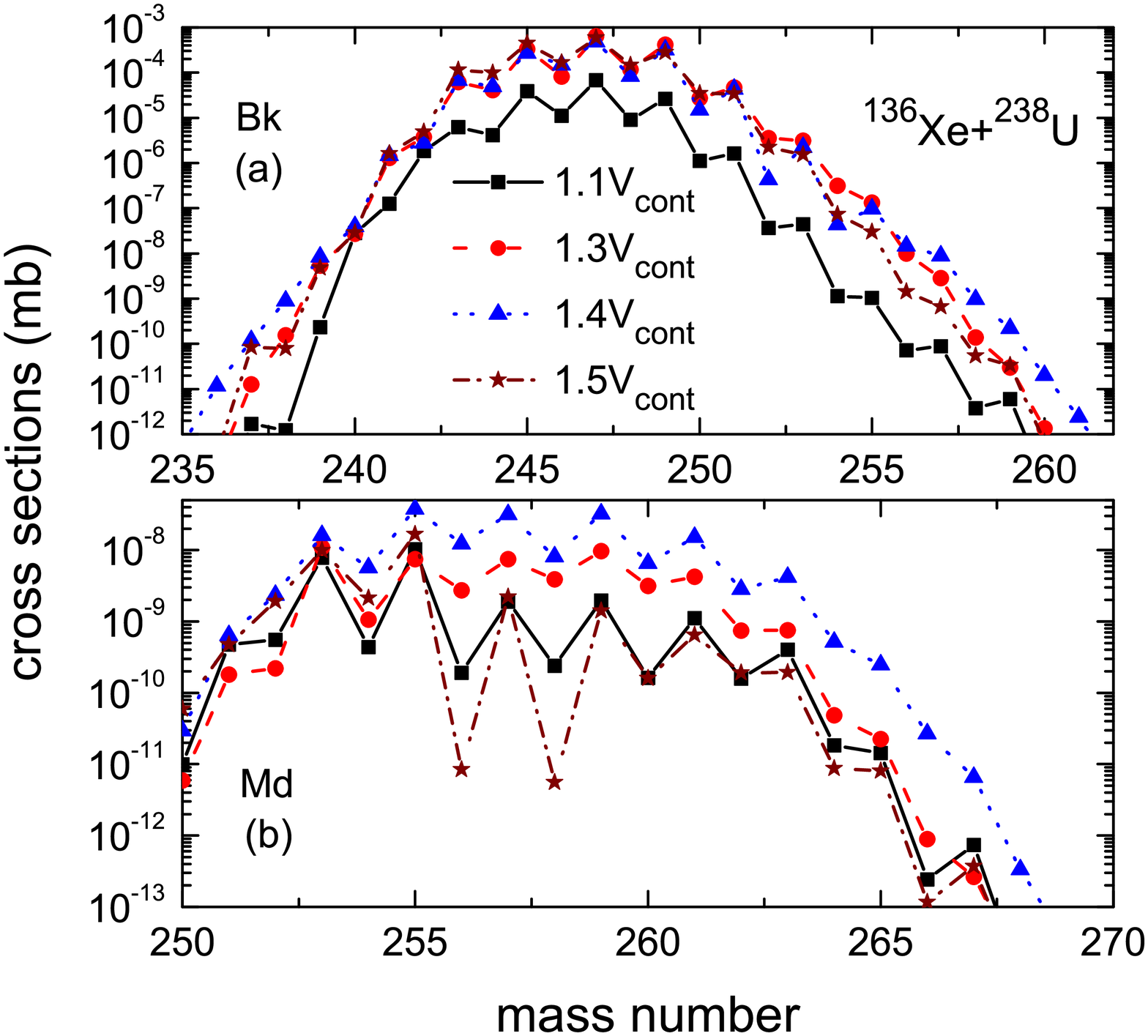}
\caption{\label{E136} Calculated cross sections of Bk (a) and Md (b) isotopes at different incident energies in the reaction $^{136}$Xe + $^{238}$U. V$_{\textrm{cont}}$=448 MeV.}
\end{center}
\end{figure}

\begin{figure}
\begin{center}
\includegraphics[width=8.5cm,angle=0]{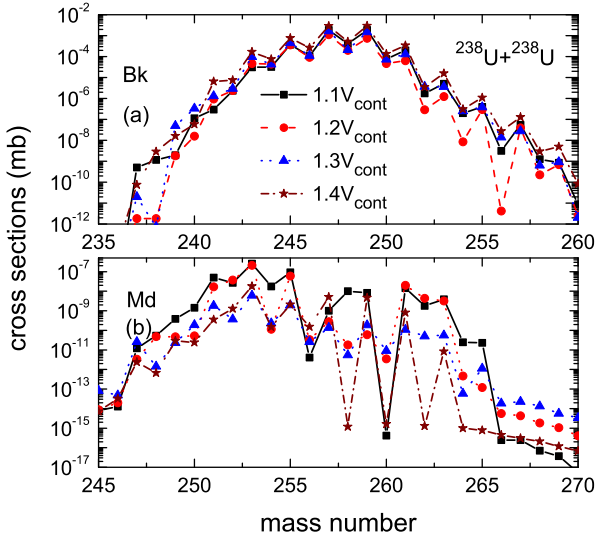}
\caption{\label{E238} Calculated cross sections of Bk (a) and Md (b) isotopes at different incident energies in the reaction $^{238}$U + $^{238}$U. V$_{\textrm{cont}}$=694 MeV.}
\end{center}
\end{figure}
It is necessary to verify the conjecture based on above analysis of PES by comparing the cross sections. As well known, the incident energy plays an important role in the MNT reactions. Therefore, in order to compare the production cross sections between different reactions, it is better to study each reaction with favorable incident energy. Figure \ref{E48} shows the production cross sections of Bk and Md isotopes in the reaction $^{48}$Ca + $^{238}$U with different incident energies. It can be seen that the production cross sections strongly depend on the incident energy, especially for producing Md isotopes. In Ref. \cite{Zhu08}, the incident energy dependence of cross sections for producing heavy nuclei around $N=126$ and actinide nuclei in the reaction $^{160}$Gd + $^{186}$W were studied. It was stated that for producing neutron-rich actinide nuclei, the final yields strongly depend on the incident energy because of high fission probability. In consideration of producing neutron-rich nuclei, the incident energy of $E_{\textrm{c.m.}}$=1.3V$_{\textrm{cont}}$, which is 220 MeV, is used as favorable one in the following study. The incident energy effects on production cross sections in the reactions $^{136}$Xe + $^{238}$U and $^{238}$U + $^{238}$U are also shown in Fig. \ref{E136} and Fig. \ref{E238}, respectively. The incident energies of $E_{\textrm{c.m.}}$=627 and 902 MeV, which are 1.4 and 1.3 times corresponding V$_{\textrm{cont}}$, are used in the following calculation for the reactions $^{136}$Xe + $^{238}$U and $^{238}$U + $^{238}$U, respectively.

\begin{figure}
\begin{center}
\includegraphics[width=8.5cm,angle=0]{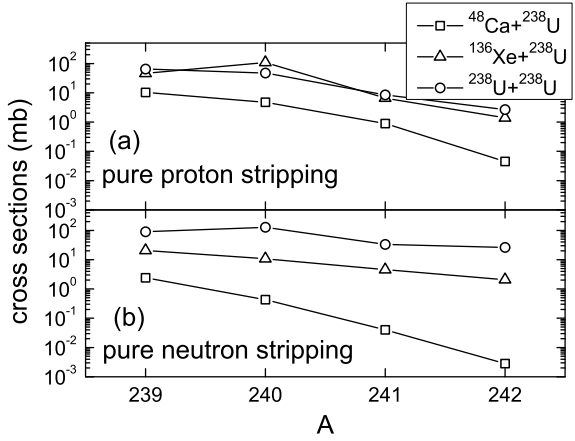}
\caption{\label{Ctr} Calculated cross sections in pure proton stripping (a) and neutron stripping (b) channels as a function of TLF mass number in the reactions $^{48}$Ca + $^{238}$U, $^{136}$Xe + $^{238}$U, and $^{238}$U + $^{238}$U. }
\end{center}
\end{figure}

\begin{figure}
\begin{center}
\includegraphics[width=8.5cm,angle=0]{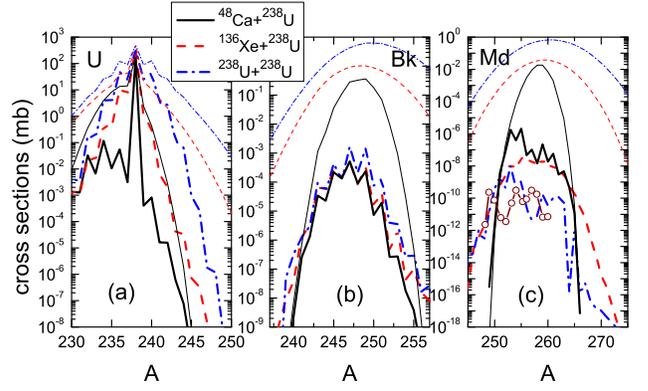}
\caption{\label{com} Comparison of calculated cross sections of U, Bk, and Md isotopes in the reactions $^{48}$Ca + $^{238}$U, $^{136}$Xe + $^{238}$U, and $^{238}$U + $^{238}$U. The thick and thin lines denote the results of final and primary products, respectively. The circles denote the results calculated by using the Langevin-type approach \cite{Saiko01}.}
\end{center}
\end{figure}

In order to testify the conjecture we made in Figs. \ref{Vp} and \ref{Vn}, it is necessary to show the cross sections of nucleon transfer in two different processes. We show in Fig. \ref{Ctr}(a) the cross sections of stripping protons in the reactions $^{48}$Ca + $^{238}$U, $^{136}$Xe + $^{238}$U, and $^{238}$U + $^{238}$U, the corresponding incident energies are E$_{\textrm{c.m.}}$=220, 627, and 902 MeV, respectively. The pure proton stripping cross sections in the reactions $^{136}$Xe + $^{238}$U and $^{238}$U + $^{238}$U are close. Because of mass asymmetry relaxation, the cross sections in the reaction $^{48}$Ca + $^{238}$U are lower than those in other reactions. The cross sections in pure neutron stripping process are also shown in Fig. \ref{Ctr}(b). As we expected, the cross sections strongly depend on the projectiles. The cross sections of transferring neutrons from $^{48}$Ca to $^{238}$U is significantly suppressed. The advantages of $^{238}$U projectile for transferring neutrons to the target is noticed.

One may wonder that whether the advantages still exist in the $^{238}$U induced reaction for the production cross sections of final products. With consideration of de-excitation, we compare the production cross sections of U, Bk, and Md isotopes among the reactions $^{48}$Ca + $^{238}$U, $^{136}$Xe + $^{238}$U, and $^{238}$U + $^{238}$U in Fig. \ref{com}. The thin lines denote the results of primary products. It is shown clearly that the distribution of U, Bk, and Md isotopes is much higher and wider in the reaction $^{238}$U + $^{238}$U than those in other ones. For producing neutron-rich uranium isotopes, the advantage of cross sections after cooling process is still huge in the reaction $^{238}$U + $^{238}$U. Unlike the uranium isotopes, for producing Bk and Md isotopes, the yields are mainly from deep inelastic collisions, which causes high excitation energies. Hence, the high fission probabilities can be seen. With increase of transferred protons, the advantages of cross sections in $^{238}$U projectile induced reaction are gradually disappear. The reasons for this behavior are as follows. (i) In proton stripping channel, the projectile effects on the cross sections of primary fragments are weak, as discussed in Fig. \ref{Ctr}(a). (ii) The excitation energies of the fragments in the reaction $^{48}$Ca + $^{238}$U are lower than those in the reaction $^{238}$U + $^{238}$U. For instance, the calculated excitation energies of $^{261}$Md at central collisions in the reactions $^{48}$Ca + $^{238}$U and $^{238}$U + $^{238}$U are 8.9 and 126 MeV, respectively. Therefore, the survival probabilities of trans-uranium products in the reaction $^{48}$Ca + $^{238}$U would be larger than those in the reaction $^{238}$U + $^{238}$U, especially for the neutron-rich isotopes. Nevertheless, due to high cross sections of primary fragments, after cooling process, the reaction $^{238}$U + $^{238}$U still shows great advantages of cross sections for producing neutron-rich Md isotopes with $A>265$. The predicted cross sections of Md isotopes by using the Langevin-type approach are also shown \cite{Saiko01}, which are close to the DNS-sysu calculation. For producing $^{253-260}$Md, the reaction $^{48}$Ca + $^{238}$U shows higher cross sections than $^{136}$Xe + $^{238}$U. However, the reaction $^{136}$Xe + $^{238}$U is more promising than $^{48}$Ca + $^{238}$U for producing neutron-rich ones. Overall, based on the $^{238}$U target, the production cross sections of neutron-rich isotopes with $Z>100$ are below the detection limit. The heavier target would be better for producing transactinide nuclei.

\begin{figure}
\begin{center}
\includegraphics[width=8.5cm,angle=0]{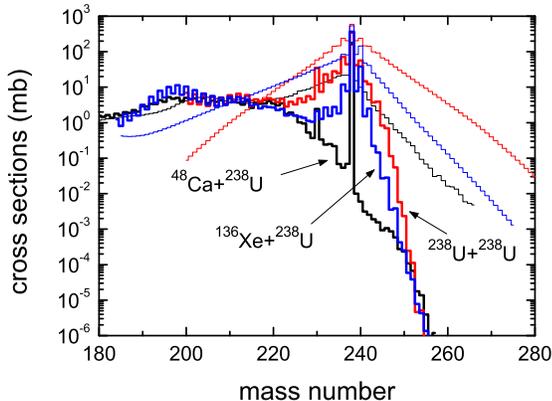}
\caption{\label{mass} Mass distributions of the products in MNT reactions $^{48}$Ca + $^{238}$U, $^{136}$Xe + $^{238}$U, and $^{238}$U + $^{238}$U. The thick and thin lines denote the results of final and primary products. }
\end{center}
\end{figure}

For further comparing the three reactions, the mass distributions of the primary MNT products in the reactions $^{48}$Ca + $^{238}$U, $^{136}$Xe + $^{238}$U, and $^{238}$U + $^{238}$U are shown in Fig. \ref{mass}. The great advantages of cross sections for producing trans-uranium nuclei can be seen in the reaction $^{238}$U + $^{238}$U. On the other side, due to mass asymmetry relaxation, the yields of primary fragments in low mass region are much higher in $^{48}$Ca induced reaction than those in other two reactions. In trans-target region, the main de-excitation channel is fission. Although with high fission probabilities, the advantage of cross sections for producing trans-target nuclei in the $^{136}$Xe and $^{238}$U induced reactions can still be noticed obviously.

\section{\label{summary}Conclusions}
The details of improvements in the DNS-sysu are presented. One statistical model based on Monte Carlo method is developed in this work. In consideration of the dynamical deformation dependent excitation energies, the descriptions of experimental cross sections are improved remarkably.

It is necessary to investigate the favorable combinations for producing unknown nuclei. The collisions of $^{48}$Ca, $^{136}$Xe, and $^{238}$U beams with $^{238}$U target are investigated within the DNS-sysu model. The influence of projectiles on cross sections of nucleon transfer are investigated. As expected from behavior of the PES, it is found that in the process of transferring neutrons from the projectile to the target, the cross sections strongly depend on the projectiles, while in pure proton stripping channels, the dependence of cross sections on projectiles is relatively weak. It is noticed that the shell effects and mass asymmetry relaxation prevent the nucleons transferring from projectiles to the target in the $^{48}$Ca and $^{136}$Xe induced reactions. The projectile $^{238}$U induced reaction show advantages of cross sections for producing neutron-rich nuclei, especially in the pure neutron transfer channel. However, with several protons transferred, the advantages are weakened. Also, it is noticed that the production cross sections of transfermium nuclei are far below the detection limit based on $^{238}$U target. For producing neutron-rich transactinide nuclei, the heavier target would be better.

\section*{ACKNOWLEDGMENTS}
The author thanks Feng-Shou Zhang, Zhao-Qing Feng, Cheng Li, Jun Su, Pei-Wei Wen, and Gen Zhang for long term discussions. This work was supported by the National Natural Science Foundation of China under Grants No. 11605296; the Natural Science Foundation of Guangdong Province, China (Grant No. 2016A030310208); and Fundamental Research Funds for the Central Universities
(18lgpy87).

\end{document}